\pdfoutput=1
\RequirePackage{ifpdf}
\ifpdf 
\documentclass[pdftex]{sigma}
\else
\documentclass{sigma}
\fi

\numberwithin{equation}{section}
\newtheorem{Theorem}{Theorem}[section]
\newtheorem*{Theorem*}{Theorem}

\newtheorem{Lemma}[Theorem]{Lemma}
\newtheorem{Proposition}[Theorem]{Proposition}
\newtheorem{Conjecture}[Theorem]{Conjecture}

\theoremstyle{definition}

\newtheorem{Example}[Theorem]{Example}
\newtheorem{Remark}[Theorem]{Remark}

\begin{document}

\allowdisplaybreaks

\newcommand{\arXivNumber}{2504.06197}

\renewcommand{\PaperNumber}{103}

\FirstPageHeading

\ShortArticleName{Orthogonal Polynomials with Complex Densities and Quantum Minimal Surfaces}

\ArticleName{Orthogonal Polynomials with Complex Densities \\ and Quantum Minimal Surfaces}

\Author{Giovanni FELDER~$^{\rm a}$ and Jens HOPPE~$^{\rm b}$}

\AuthorNameForHeading{G.~Felder and J.~Hoppe}

\Address{$^{\rm a)}$~Department of Mathematics, ETH Zurich, 8092 Zurich, Switzerland}
\EmailD{\mail{giovanni.felder@math.ethz.ch}}

\Address{$^{\rm b)}$~Technische Universit\"at Braunschweig, Germany}
\EmailD{\mail{jens.r.hoppe@gmail.com}}

\ArticleDates{Received September 01, 2025, in final form November 26, 2025; Published online December 07, 2025}

\Abstract{We show that the discrete Painlev\'e-type equations arising from quantum minimal surfaces are equations for recurrence coefficients of orthogonal polynomials for indefinite hermitian products. As a consequence, we obtain an explicit formula for the initial conditions leading to positive solutions.}


\Classification{33C45; 34M55; 53A10; 15B52}

\section{Introduction}
A quantum minimal surface \cite{Hoppe1982,Hoppe1996,IKKT1997} in $\mathbb R^N$ is described by hermitian operators $X_i$ $(i=1,\dots,N)$ satisfying the double commutator equations
\[
\sum_{i=1}^N[X_i,[X_i,X_j]]=0, \qquad j=1,\dots,N.
\]
These equations are a quantization of the equations \smash{$\sum_{i=1}^N\{x_i,\{x_i,x_j\}\}=0$} describing a minimal embedding $x\colon\Sigma\to \mathbb R^N$ of a Riemann surface $\Sigma$ with Poisson bracket $\{\ ,\ \}$ defined by the area form. If $N=2m$ is even, the double commutator equations are implied by the equations~${\sum_{j=1}^m\big[W_j^\dagger,W_j\big]=\epsilon 1}$, $[W_j,W_k]=0$ for $W_j=X_{2j-1}+{\rm i}X_{2j}$, $W_j^\dagger=X_{2j-1}-{\rm i}X_{2j}$, ($j=1,\dots,m$). In the classical case these are first order equations implying the second order minimal surface equation. Cornalba and Taylor \cite{CornalbaTaylor1999} viewed them as stationary solutions of the membrane matrix model \cite{Hoppe1982}. For $N=4$, they considered solutions where $W_2=X_3 + {\rm i} X_4$ is the square of~${W_1=X_1 +{\rm i} X_2}$, for which an interesting recurrence relation
\begin{equation}\label{e-00}
v_n(1+v_{n+1}+v_{n-1})=(n+1)\epsilon,
\end{equation}
with a parameter $\epsilon>0$ arises for the square norms of the matrix elements of $W$ (assumed to be non-zero only on one of the first off-diagonals). In the classical limit, the corresponding minimal surface is a special case of the class of minimal surfaces given by a holomorphic relation~${f(x_1+{\rm i}x_2,x_3+{\rm i}x_4)=0}$ discovered over a century ago \cite{Eisenhart1912,Kommerell1905}. The recurrence relation~\eqref{e-00} is a type of discrete Painlev\'e equation: it is called ternary dP${}_{\mathrm{I}}$ in~\cite{Tokihiro2002}, where it is shown to be a~contiguity relation for the Painlev\'e differential equation P${}_{\mathrm V}$.
In \cite{AHK2019}, apart from discussing several other examples and various aspects of quantum minimal surfaces, such as Hermitian Yang--Mills theory (see also \cite{Connes2002, Herscovich2011} as well as, related in yet other ways, \cite{ACH2016,AHH2012}), several properties of the recurrence relations were given, while in \cite{CDHM2025,Hoppe2025}
the unique positive solution was described.

Another source of recurrence relations such as \eqref{e-00} is the theory of random matrices and orthogonal polynomials, where they arise as equations for recurrence coefficients. Indeed, another discrete Painlev\'e equation, called dP{}$_{\mathrm I}$, namely
$v_n(1+v_{n+1}+v_n+v_{n-1})=(n+1)\epsilon$,
occurs in hermitian random matrices with quartic potentials and is related to P${}_{\mathrm{IV}}$, see the review \cite{Its2011} and references therein. Both dP{}$_{\mathrm I}$ and the ternary dP{}$_{\mathrm I}$ have as a continuum limit the differential Painlev\'e I equation.

In this paper, we develop a theory of orthogonal polynomials leading to \eqref{e-00} and other equations arising in quantum minimal surfaces. The starting point of our research was the observation that in the theory of random {\em normal} matrices \cite{WiegmannZabrodin2000} a recurrence relation appears \cite{TBAZW2005} that differs from \eqref{e-00} only by a sign change,
see also \cite{AptekarevNovokshenov} for a recent analysis of the asymptotics of solutions. The orthogonal polynomials are defined by a density on the complex plane. To get the right sign, we consider a variant with complex density which seems to be interesting in its own right. For example, the recurrence relation \eqref{e-00} arises if we consider an inner product on polynomials given by an integral over the complex plane
\begin{equation}\label{e-ip}
(f,g)=\int_{\mathbb C}f(z)\overline{g(-z)}{\rm e}^{{-a|z|^2+{\rm i} (V(z)+\overline{V(z)})}}{\rm d}x{\rm d}y,\qquad f,g\in\mathbb C[z],
\end{equation}
with cubic potential $V(z)=tz^3$. The parameters $a$, $t$ are real with $a>0$. The integral is over~${z=x+{\rm i}y\in\mathbb C}$ and converges absolutely, in contrast to the integrals occurring in normal random matrix models (with $t$ imaginary in this example) \cite{Elbau2007,TBAZW2005,WiegmannZabrodin2000} which require a regularization, see \cite{BleherKuijlaars2012,ElbauFelder2005}.
The downside is that the integration density is not real making the probabilistic interpretation questionable.
Still, this inner product is hermitian
$
(f,g)=\overline{(g,f)} $ for all $ f,g\in\mathbb C[z]$,
and in particular $(f,f)$ is real for all polynomials $f$.
The inner product is not definite in general, so it is not guaranteed that orthogonal polynomials exist. If they exist (and they do in the case of a cubic potential as follows from \cite{CDHM2025,Hoppe2025} as we show in Theorem~\ref{t-3}), monic orthogonal polynomials are unique and produce positive solutions of \eqref{e-00} and thus a quantum minimal surface. Moreover, the theory produces the explicit initial condition that leads to the positive solution of~\eqref{e-00}.

Another hint that minimal surfaces are related to normal random matrices comes from the observation that the eigenvalue distribution of regularized normal random matrices \cite{ElbauFelder2005,WiegmannZabrodin2000} with cubic potential are uniformly distributed on a domain in the complex plane bounded by a~hypotrochoid. This is one of the special shapes that when rotated give rise to a minimal surface in Minkowski space \cite{Hoppe2009}. We do not pursue this relation in this paper, but plan to come back to it in the future.

In Section \ref{s-2} of this paper, we discuss the case of a cubic potential in details and explain how the relation \eqref{e-00} arises from the coefficients of the recurrence relation for orthogonal polynomials. The initial condition is expressed as a modified Bessel function as in \cite{CDHM2025,Hoppe2025}.

In Section \ref{s-3}, we generalize the construction to the case of monomial potentials, replacing~$z^3$ by $z^d$ for $d\in\mathbb Z_{>0}$. The corresponding quantum minimal surfaces are solutions $W_1$, $W_2$ such that $W_2=W_1^{d-1}$. The existence of ``integrable'' recurrence relations generalizing \eqref{e-00} is mentioned in \cite{AHK2019} and was written explicitly in \cite{CDHM2025,Hoppe2025} for $d=4$. It turns out that for odd $d$ the same formula~\eqref{e-ip} with $3$ replaced by $d$ gives a hermitian inner product and everything generalizes nicely. For even $d$, a modification is needed, related to the fact that the extension group Ext($\mathbb Z/d\mathbb Z,\mathbb Z/2\mathbb Z)$ is non-trivial for $d$ even. Once this is understood, we get an indefinite hermitian inner product on polynomials of degree $\leq N$ for all~$N$. We conjecture that it is non-degenerate for all $N$ and all values of the parameters. This conjecture implies the existence of orthogonal polynomials and of a positive solution of the recurrence relation.

This solution can be computed explicitly as we show in Section~\ref{s-4}. We discuss the initial conditions of the recurrence relation produced by orthogonal polynomials. They are determined by a single function $h(a)$ (the square norm of the polynomial 1) which obeys a linear differential equation of order $d-1$, whose solutions are generalized hypergeometric functions. We give explicit formulas for the coefficients and express $h$ in terms of classical functions for small $d$.

In Section \ref{s-5}, we explain the relation to minimal surfaces, and in Section \ref{s-6}, we discuss some future research directions.

\section{Cubic potential}\label{s-2}
Let $a>0$, $t\in\mathbb R$. Consider the sesquilinear pairing
\begin{equation}\label{e-1}
(f,g)=\int_{\mathbb C}f(z)\overline{g(-z)}\rho_{a,t}(z){\rm d}x{\rm d}y,\qquad
\rho_{a,t}(z)=\exp\bigl(-a|z|^2+{\rm i}t\bigl(z^3+\bar z^3\bigr)\bigr),
\end{equation}
on polynomials $f,g\in\mathbb C[z]$. The integral is over $z=x+{\rm i}y\in\mathbb C$ and converges absolutely.

Let us assume that $(\ ,\ )$ is non-degenerate.
Polynomials $P_0,P_1,\dots$ are called orthogonal for~$(\ ,\ )$ if
\begin{enumerate}\itemsep=0pt
 \item[(1)] $P_n(z)=z^n+\cdots$ is monic of degree $n$,
 \item[(2)]$(P_n,P_m)=0$ if $n\neq m$.
\end{enumerate}
It is not a priori clear that orthogonal polynomials exist, since the inner product is not definite in general, but if they exist and the inner product is non-degenerate, they are unique. In~fact, it~is a consequence of our construction, together with existence of the unique positive solution to~\eqref{e-00} that orthogonal polynomials $P_n$ exist for all $n$ and that the sign of $(P_n,P_n)$ is $(-1)^n$, see Theorem~\ref{t-3} below.

If $(P_n)_{n\geq0}$ is the sequence of orthogonal polynomials, we set
$
h_n=(-1)^n(P_n,P_n)$.
The sign is chosen so that $h_n>0$ for $t=0$:
\begin{Example}
 Let $t=0$. Then $P_n(z)=z^n$ are orthogonal and
 \[
 h_n=\int_{\mathbb C} |z|^{2n}{\rm e}^{-a|z|^2}{\rm d}x{\rm d}y=\frac{\pi n!}{a^{n+1}}.
 \]
\end{Example}
We see that for $t=0$ the restriction to polynomials of degree $\leq N$ has signature $(k,k)$ if~${N=2k-1}$ is odd and $(k+1,k)$ if $N=2k$ is even. This continues to be the case for small $t$ by continuity and also for larger $t$ as long as the inner product stays non-degenerate. We show in the next section that the inner product is non-degenerate for all $a>0$, $t\in\mathbb R$.

\subsection{Action of the cyclic group}
Let $\zeta=\exp(2\pi {\rm i}/3)$, $T$ the automorphism of the ring $\mathbb C[z]$ such that $z\mapsto \zeta z$. Then $(Tf,Tg)=(f,g)$ for all $f,g$ so that $T$ defines an action of the cyclic group $C_3$ of order 3 on polynomials. By the uniqueness of orthogonal polynomials, we see that
$TP_n=\zeta^nP_n$.

\subsection{Recurrence relation for orthogonal polynomials}
Integration by parts leads to the identity
\begin{equation}\label{e-0}
\bigl(\partial_zf+3{\rm i}tz^2f,g\bigr)=-a(f,zg).
\end{equation}
The following statement shows that we can effectively recursively determine the orthogonal polynomials if we know the $h_n$.
\begin{Proposition}\label{p-1} Let $L=\partial_z+3{\rm i}tz^2$. Then $P_0(z)=1$, $P_1(z)=z$, $P_2(z)=z^2$ and
 \begin{gather*}
 zP_n(z) = P_{n+1}(z)+\frac{3{\rm i}t}{a}\frac{h_n}{h_{n-2}}P_{n-2}(z),\qquad n\geq2,
 \\
 LP_n(z) = 3{\rm i}t P_{n+2}(z)+a\frac{h_n}{h_{n-1}}P_{n-1}(z),\qquad n\geq1.
 \end{gather*}
\end{Proposition}
\begin{proof}
 $1$, $z$, $z^2$ are the only monic polynomials of degree $\leq2$ of the correct $C_3$-weight.

 The polynomial $zP_n$ has degree $n+1$ and is therefore a linear combination of $P_j$ with $j\leq n+1$. Similarly, $LP_n$ is a linear combination of $P_j$ with $j\leq n+2$. Thus $(zP_n,P_m)=-a^{-1}(P_n,LP_m)$ vanishes for $m<n-2$ and $zP_n$ is a linear combination of $P_{n+1}$, $P_n$, $P_{n-1}$, $P_{n-2}$. Since $z$ has weight 1, for the $C_3$-action
 $ zP_n=a_nP_{n+1}+b_nP_{n-2}$
 for some coefficients $a_n$, $b_n$.
 By the condition that $P_n$ is monic, we obtain $a_n=1$. Similarly, the action of the operator $L$ of weight 2 has the form
 $
 LP_n=3{\rm i}t P_{n+2}+c_nP_{n-1}$.
 The coefficient $b_n$ is determined by the relation
 \[
 ab_n(P_{n-2},P_{n-2})=a(zP_n,P_{n-2})=-(P_n,LP_{n-2})=-(P_n,3{\rm i}t P_n)=3{\rm i}t(P_n,P_n).
 \]
 For $c_n$, we obtain
 \[
 c_n(P_{n-1},P_{n-1})=(LP_n,P_{n-1})=-a(P_n,zP_{n-1})=-a(P_n,P_n).\tag*{\qed}
 \]
 \renewcommand{\qed}{}
\end{proof}

\begin{Remark}\label{r-1}
 Let $V_n=h_{n+1}/h_n$, $n\geq0$. Then we can write the formulas of Proposition~\ref{p-1}~as
 \begin{align*}
 zP_n(z) = P_{n+1}(z)+\frac{3{\rm i}t}{a}V_{n-1}V_{n-2}P_{n-2}(z),
 \qquad
 LP_n(z) = 3{\rm i}t P_{n+2}(z)+aV_{n-1}P_{n-1}(z).
 \end{align*}
 We notice that since $P_n=z^n$ for $n=0,1,2$ these formulas hold for all $n\geq0$, provided we set~${V_{n}=0}$ for $n<0$.
\end{Remark}

\subsection{A discrete Painlev\'e equation}
\begin{Theorem}\label{p-2}
 The ratios $V_n=h_{n+1}/{h_n}$, $n\geq0$, obey the recurrence relation
 \[
 V_n\left(1+\frac{9t^2}{a^{2}}(V_{n-1}+V_{n+1})\right)=\frac{n+1}a
 \]
 with initial conditions
 \[
 V_{-1}=0,\qquad V_0=\frac{\int_{\mathbb C}|z|^2\rho_{a,t}(z){\rm d}x{\rm d}y}{\int_{\mathbb C}\rho_{a,t}(z){\rm d}x{\rm d}y}.
 \]
\end{Theorem}
\begin{proof}
 Proposition~\ref{p-1} gives the matrix elements of multiplication by $z$ and of $L=\partial_z+3{\rm i}t z^2$ in the basis of orthogonal polynomials. From it, we can deduce the action of $\partial_z$. The condition that the coefficient of $P_{n-1}$ in the expression for $\partial_z P_n$ must be $n$ gives a condition for the numbers~$h_n$. The calculation goes as follows. By Remark \ref{r-1},
 \begin{align*}
 z^2P_n&=z\left(P_{n+1}+3{\rm i}\frac taV_{n-1}V_{n-2}P_{n-2}\right)
 \\
 &=P_{n+2}+3{\rm i}\frac taV_{n}V_{n-1}P_{n-1}
 +3{\rm i}\frac taV_{n-1}V_{n-2}\left(P_{n-1}+3{\rm i}\frac taV_{n-3}V_{n-4}P_{n-4}\right)
 \\
 &=P_{n+2}+3{\rm i}\frac taV_{n-1}(V_n+V_{n-2})P_{n-1}-(\cdots)P_{n-4},
 \end{align*}
 (we do not care what the coefficient of $P_{n-4}$ is). Thus
 \begin{align*}
 \partial_zP_n&=LP_n-3{\rm i}tz^2P_n
 \\
 &=3{\rm i}tP_{n+2}(z)+aV_{n-1}P_{n-1}
 -3{\rm i}t\left(P_{n+2}+3{\rm i}\frac ta V_{n-1}(V_n+V_{n-2})P_{n-1}+(\cdots)P_{n-4}\right)\\
 &=\left(aV_{n-1}+9\frac{t^2}aV_{n-1}(V_{n}+V_{n-2})\right)P_{n-1}+(\cdots)P_{n-4}.
 \end{align*}
 Equating the coefficient of $P_{n-1}$ with $n$ yields the recurrence relation.
\end{proof}

 Let
 $ v_n=\frac{9t^2}{a^2}V_n$, $ \epsilon = \frac{9t^2}{a^3}$.
 Then we can write the recurrence relation as $v_n(1+v_{n+1}+v_{n-1})=(n+1)\epsilon$.
The initial conditions are $v_{-1}=0$ and a function $v_0$ of $\epsilon$.
This function can be expressed in terms of Bessel functions
\smash{$
v_0(\epsilon)=\epsilon^{\frac23}f\bigl(\epsilon^{-\frac13}\bigr)$},
where
\[
f(a)=-h'(a)/h(a),\qquad h(a)=\int_{\mathbb C}\exp\left(-a|z|^2+\frac{\rm i}3\bigl(z^3+\bar z^3\bigr)\right){\rm d}x{\rm d}y.
\]
\begin{Lemma}\label{l-1}
The function $h(a)$ is the solution of the differential equation
$h''(a)=ah(a)+a^2h'(a)$
with the boundary condition $\lim_{a\to\infty}ah(a)=\pi$.
\end{Lemma}
\begin{proof}
 We differentiate under the integral and integrate by parts
 \begin{align*}
 h''(a)&=\int_{\mathbb C} z^2\bar z^2 {\rm e}^{-a|z|^2+\frac{\rm i}3(z^3+\bar z^3)}{\rm d}x{\rm d}y
 =-\int_{\mathbb C} {\rm e}^{-az\bar z}\partial_z\partial_{\bar z}{\rm e}^{\frac{\rm i}3(z^3+\bar z^3)}{\rm d}x{\rm d}y
 \\
 &=-\int_{\mathbb C}\partial_z \partial_{\bar z}{\rm e}^{-az\bar z}{\rm e}^{\frac{\rm i}3(z^3+\bar z^3)}{\rm d}x{\rm d}y
 =-\int_{\mathbb C}\bigl(-a+a^2z\bar z\bigr){\rm e}^{-az\bar z}{\rm e}^{\frac{\rm i}3(z^3+\bar z^3)}{\rm d}x{\rm d}y
 \\
 &=ah(a)+a^2h'(a).
 \end{align*}
 To compute the behavior at infinity, we rescale $z$ by $a^{-\frac12}$
 \[
 h(a)=a^{-1}\int_{\mathbb C} \exp\left({-|z|^2+\frac{\rm i}3a^{-\frac32}}\bigl(z^3+\bar z^3\bigr)\right){\rm d}x{\rm d}y.
 \]
 Thus
 \[
 \lim_{a\to\infty}ah(a)=\int_{\mathbb C} {\rm e}^{-|z|^2}{\rm d}x{\rm d}y=\pi.\tag*{\qed}
\]
 \renewcommand{\qed}{}
\end{proof}

The change of variables
\smash{$
h(a)={\rm e}^{a^3/6}\sqrt{a} y\bigl(a^3/6\bigr)$} leads to the modified Bessel differential equation
\[
x^2\frac{{\rm d}^2y}{{\rm d}x^2}+x\frac{{\rm d}y}{{\rm d}x}-\left(x^2+\frac1{36}\right)y=0.
\]
The solutions vanishing at infinity are proportional to the modified Bessel function of the second kind $K_{1/6}(x)$ behaving asymptotically as $\sqrt{\frac\pi{2x}}{\rm e}^{-x}(1+O(1/x))$ for $x\to\infty$. We conclude that
\begin{equation}\label{e-K}
h(a)={\rm e}^{a^3/6}\sqrt{\frac{\pi a}3}K_{1/6}\bigl(a^3/6\bigr).
\end{equation}
\section{Orthogonal polynomials with complex density}\label{s-3}
We wish to generalize the above construction to the case where the exponent 3 is replaced by an arbitrary natural number $d$. This is straightforward if $d$ is odd, but for even $d$ a more general construction is needed. Let $d\in\mathbb Z_{\geq0}$ and $\mu_d=\big\{\zeta\in\mathbb C \mid \zeta^d=1\big\}$ be the group of $d$-th roots of unity acting on $\mathbb C$ by multiplication. Suppose $\rho\colon\mathbb C\to \mathbb C$ is a measurable complex valued function such that
\begin{enumerate}\itemsep=0pt
 \item [(I)] $\int_{\mathbb C}|z|^n|\rho(z)|{\rm d}x{\rm d}y<\infty$ for all $n\in\mathbb Z_{\geq0}$,
 \item [(II)] $\rho(\eta z)=\overline{\rho(z)}$ for all $\eta\in \mathbb C$ such that $\eta^d=-1$.
\end{enumerate}
Note that (II) implies that $\rho$ is invariant under $\mu_d$, $\rho(\zeta z)=\rho(z)$ for all $\zeta\in\mu_d$. Also any two solutions of $\eta^d=-1$ differ by multiplication by an element of $\mu_d$. So one can replace (II) by
\begin{enumerate}\itemsep=0pt
 \item [(II$'$)] $\rho$ is $\mu_d$-invariant and $\rho(\eta z)=\overline{\rho(z)}$ for some $\eta$ such that $\eta^d=-1$.
\end{enumerate}
Property (I) implies that the integrals defining the moments $\int_{\mathbb C} z^n\bar z^m\rho(z){\rm d}x{\rm d}y$ are absolutely convergent. The group $\mu_d$ acts
on polynomials via $T_\zeta f(z)=f(\zeta z)$, $\zeta\in\mu_d$.
The ring of polynomials $\mathbb C[z]$ decomposes into a direct sum
\smash{$
\mathbb C[z]=\bigoplus_{\ell\in\mathbb Z/d\mathbb Z}\mathbb C[z]_\ell
$}
of weight spaces
\[
\mathbb C[z]_\ell=\big\{f \mid T_\zeta f=\zeta^\ell f
\text{ for all $\zeta\in\mu_d$}\big\}.
\]
To a function $\rho$ as above, we associate a hermitian $\mu_d$-invariant inner product. The construction depends on the choice of a right inverse $\chi$ of the natural projection $\mathbb Z/2d\mathbb Z\to\mathbb Z/d\mathbb Z$. For definiteness, we take
$\chi(\ell \bmod{d})=\ell \bmod{2d}$ for $\ell=0,\dots,d-1$.

\begin{Proposition}
 Let $\rho$ be a function obeying {\rm (I)} and {\rm (II)} for $d\in\mathbb Z_{\geq1}$ and $\eta\in\mathbb C$ such that~${\eta^d=-1}$.
 Let $\sigma\colon \mathbb C[z]\to\mathbb C[z]$ the linear map such that $\sigma f(z)=(-\eta)^{-\chi(\ell)} f(\eta z)$ if $f\in\mathbb C[z]_\ell$. Then the inner product
 \[
 (f,g)=\int_{\mathbb C}f(z)\overline{\sigma g(z)}\rho(z){\rm d}x{\rm d}y
 \]
 is independent of the choice of $\eta$ and hermitian. In particular, $(f,f)\in\mathbb R$ for all $f\in\mathbb C[z]$.
 Moreover, $(T_\zeta f,T_\zeta g)=(f,g)$ for $\zeta\in\mu_d$ and any polynomials $f$, $g$.
\end{Proposition}
\begin{proof}
 Any two choices of $\eta$ differ by multiplication by a root of unity $\zeta\in\mu_d$. But $\sigma$ does not change if we replace $\eta$ by $\eta\zeta$, so the inner product is unchanged.

 Also $\sigma$ commutes with the action of $\mu_d$, $\rho$ and the measure ${\rm d}x{\rm d}y$ are $\mu_d$-invariant. implying that last claim. In particular, $(f,g)=0$ if $f$ and $g$ have different weight.
 To check the hermitian property, we can thus assume that $f$ and $g$ have the same weight $\ell$.
 We use (II) and that $\eta^2\in\mu_d$,
 \begin{align*}
 \overline{(g,f)}&=\int_{\mathbb C}\overline{g(z)}(-\eta)^{-\chi(\ell)}f(\eta z)\rho(\eta z){\rm d}x{\rm d}y
 =\int_{\mathbb C}\overline{g\big(\eta^{-1} z\big)}(-\eta)^{-\chi(\ell)}f(z)\rho(z){\rm d}x{\rm d}y
 \\
 &=\int_{\mathbb C}\overline{g\big(\eta^{-2}\eta z\big)}(-\eta)^{-\chi(\ell)}f(z)\rho(z){\rm d}x{\rm d}y
 =\int_{\mathbb C}\overline{\eta^{-2\ell}g(\eta z)}(-\eta)^{-\chi(\ell)}f(z)\rho(z){\rm d}x{\rm d}y
 \\
 &=\int_{\mathbb C}\overline{(-1)^{-\chi(\ell)}\eta^{-2\ell+\chi(\ell)}g(\eta z)}f(z)\rho(z){\rm d}x{\rm d}y
 \\
 &=\int_{\mathbb C}\overline{(-\eta)^{-\chi(\ell)}g(\eta z)}f(z)\rho(z){\rm d}x{\rm d}y=(f,g).
 \end{align*}
 In the last step, we used the fact that $2\ell\equiv 2\chi(\ell)\bmod 2d$ so that $\eta^{-2\ell}=\eta^{-2\chi(\ell)}$
 for $\eta\in\mu_{2d}$.
\end{proof}

\begin{Remark}
 To construct the inner product, we chose a right inverse of the natural $2:1$ projection $p\colon \mathbb Z/2d\mathbb Z\to \mathbb Z/d\mathbb Z$. Any of the other $2^d$ choices would also lead to a hermitian inner product. In fact, for any map $\lambda\colon\mathbb Z/d\mathbb Z\to\{\pm1\}$, the inner product $(\ ,\ )_s$ whose restriction to the weight space $\mathbb C[z]_\ell$ is
 $
 (f,g)_\lambda=\lambda(\ell)(f,g)$, if $f\in \mathbb C[z]_\ell$
 is also a $\mu_d$ invariant hermitian inner product.
\end{Remark}

Here is the class of examples we have in mind.

\begin{Example}\label{ex-1}
 Let $\rho(z)=\exp\bigl(-a|z|^2+{\rm i}\bigl(V(z)+\overline {V(z)}\bigr)\bigr)$, where $V(z)=U\bigl(z^d\bigr)$ for some {\em odd} entire function $U$. If $d$ is odd, we can take $\eta =-1$ and the inner product is given by the same formula \eqref{e-1}.
\end{Example}
The integration by part formula \eqref{e-0} for this class of examples generalizes, but in the case of even $d$ there is a twist by a sign.

\begin{Lemma}
 Let $\rho$ be as in Example~{\rm\ref{ex-1}} with polynomial $V$ and set $L=\partial_z + {\rm i}V'(z)$.
 Let $\tau\colon\mathbb Z/d\mathbb Z\to\{\pm1\}$ be such that
 \[
 \tau(\ell)=\begin{cases}
 (-1)^{d-1}& \text{if $\ell\equiv -1\mod{d}$},
 \\
 \phantom{-}1& \text{otherwise.}
 \end{cases}
 \]
 Then
 $ (Lf,g)=-\tau(\ell)a\cdot(f,zg)$ for any polynomials $f$, $g$ such that $g$ has weight $\ell$,
\end{Lemma}
Again, orthogonal polynomials are not guaranteed to exist. A sufficient condition for existence of monic orthogonal polynomials $P_0,\dots, P_N$ up to degree $N$ is the non-vanishing of the determinants of the Gram matrices up to degree $N$. The $N$-th Gram matrix of the inner product is the matrix $G_N$ with entries $(z^n,z^m)$, $n,m=0,\dots, N$. If all Gram matrices $G_j$ for $j$ up to $N$ are invertible, then $(\ ,\ )$ admits orthogonal polynomials $P_j=z^j+\cdots$ up to degree $N$ and they are unique.

\begin{Example}\label{ex-2}
 Let $\rho(z)=\exp\bigl(-a\big|z^2\big|\bigr)$. Then $1,z,z^2,\dots$ are orthogonal polynomials for any $d$ and
 \[
 (z^n,z^n)=\rho(n)\frac{\pi n!}{a^{n+1}},
 \]
 where $\rho(n)=(-1)^n\prod_{j=0}^{n-1}\tau(j)$. Explicitly, $\rho(n)=(-1)^n$ for $n=0,\dots,d-1$ and $\rho(n+d)=-\rho(n)$ for general $n$.
\end{Example}
By continuity of the determinant of the Gram matrices, it is still true that orthogonal polynomials $P_0,\dots, P_N$ exist for $\rho$ as in Example~\ref{ex-1} for any fixed $N$, provided $V$ is a polynomial with sufficiently small coefficients (depending on $N$). The sign of $(P_n,P_n)$ is then still $\rho(n)$. It is an interesting question to find criteria for the potential $U(z)$ ensuring the non-degeneracy of the Gram matrices and thus the existence of orthogonal polynomials.

\subsection{Monomial potentials}
In the same way as in the case of cubic monomial potentials, we can consider the case of monomials of general degree $d$. The density of the inner product \eqref{e-1} is replaced by
\[
\rho_{a}(z)=\exp\left(-a|z|^2+\frac {\rm i}d\bigl(z^d+\bar z^d\bigr)\right),
\]
where $d$ is a positive integer, $a>0$.\footnote{For simplicity, we set $t=1/d$. The case with a general $t$ can be recovered by rescaling variables.} Then the inner product is hermitian and we can define orthogonal monic polynomials $P_n$ with square norms $(P_n,P_n)=\rho(n)h_n$.
\begin{Conjecture}\label{con}
 The Gram matrices $((z^n,z^m))_{n,m=0}^N$, $N=0,1,2,\dots$, are invertible for all $ a\geq0$.
\end{Conjecture}
For small $d$, the conjecture holds as a consequence of the results of \cite{CDHM2025,Hoppe2025}.
\begin{Theorem}\label{t-3}
The conjecture holds for $d=1,2,3$.
\end{Theorem}

\begin{proof} Let us fix the maximal degree $N$. Then the inner product multiplied by a power of~$a$ converges for $a\to\infty$ to the inner product with density $\exp\bigl(-|z|^2\bigr)$ for which the $z^n$ are orthogonal with non-vanishing squared norms $\rho(n)\pi n!$. Thus the Gram matrices up to $N$ are non-degenerate for large $a$, orthogonal polynomials exist, and the determinant of the Gram matrix is (up to sign) the product $\prod_{n=0}^Nh_n(a)$. Assume by contradiction that for some $a>0$ a~Gram matrix is degenerate and let $a_*$ be the largest $a$ for which this happens. This means that $h_{n+1}(a_*)=0$ for some $n\geq0$, which can be chosen as small as possible so that
\begin{gather*}
h_{j}(a)>0 \qquad\text{for all}\ a\geq a_* \ \text{and}\ j\leq n,
\\
h_{n+1}(a)>0\qquad \text{for all}\ a>a_*,
 \\
 h_{n+1}(a_*)=0.
\end{gather*}
This means that there is an $n\geq1$ and $a_*>0$ such that $V_n(a)$ vanishes at $a_*$ and is positive for~${a>a_*}$ while $V_0(a),\dots,V_{n-1}(a)$ are positive for all $a\geq a_*$. For $d=1,2$, the $V_n$ can be computed explicitly, see below, and they are positive for all $a$. For $d=3$, it is shown in \cite{CDHM2025,Hoppe2025} that the initial condition $V_0(a)=-h'(a)/h(a)$ with $h$ given by \eqref{e-K} gives rise to a solution for which all $V_n(a)>0$ for all $a>0$, so no such $a_*$ can exist.
We reproduce here the argument of~\cite{Hoppe2025} to prove this in the language of orthogonal polynomials. It relies on a first order differential equation for $V_n(a)$ which can be obtained by observing that the derivative of $(P_n,P_n)$ with respect to $a$ is proportional to $(zP_n,zP_n)$ which can be expressed in terms of $(P_{n+1},P_{n+1})$ and~${(P_{n-2},P_{n-2})}$ by using the relation \eqref{e-rec1}. The result is
\[
\partial_ah_n=-h_{n+1}+\frac1{a^2}\frac{h_n^2}{h_{n-d+1}},
\]
where we take $d=3$. This translates to a differential equation for $V_n=h_{n+1}/h_n$
\[
\partial_aV_n=-V_nV_{n+1} +V_n\left(V_n+\frac1{a^2}(V_n-V_{n-2})V_{n-1}\right).
\]
Now we can eliminate $V_{n+1}$ using the recurrence relation $V_nV_{n+1}=-V_nV_{n-1}-a^2V_n+(n+1)a$ and obtain a differential equation of the form
\[
\partial_aV_n(a)=-(n+1)a+V_n(a) P(V_{n-2}(a),V_{n-1}(a),V_n(a))
\]
for a polynomial $P$. The point is that at $a=a_*$, where $V_n$ vanishes, the derivative $\partial_aV_n(a_*)=-(n+1)a_*$ is negative, in contradiction with $V_n(a)$ being positive for $a>a_*$.
\end{proof}

In the following, we assume the validity of this conjecture for general $d$. It implies that orthogonal polynomials exist and that $h_n>0$ for all $n$.

The polynomials $P_n$ have weight $n$ under the unitary action of $\mu_d$, meaning that $P_n(\zeta z)=\zeta^nP_n(z)$ for $d$-th roots of unity $\zeta$. The first $d$ polynomials are $1,\dots,z^{d-1}$.
The relations of Proposition~\ref{p-1} become
\begin{gather}\label{e-rec1}
 zP_n(z) = P_{n+1}(z)+\frac{\rm i}{a}\frac{h_n}{h_{n-d+1}}P_{n-d+1}(z),\qquad n\geq d-1,
 \\
 LP_n(z) = {\rm i} P_{n+d-1}(z)+a\frac{h_n}{h_{n-1}}P_{n-1}(z),\qquad n\geq1, \label{e-rec2}
\end{gather}
where $L=\partial_z+{\rm i}z^{d-1}$.
By the calculation of Theorem~\ref{p-2}, the ratios $V_n=h_{n+1}/h_n$ obey the recurrence relations
\begin{equation}\label{e-3}
V_n+\frac{1}{a^2}\sum_{j=0}^{d-2}\prod_{k=0}^{d-2}V_{n+j-k}=\frac{n+1}a,
\end{equation}
with initial conditions $V_{-1}=0$ and
\[
V_j=\frac{\int_{\mathbb C}|z|^{2j+2}\rho_{a}(z){\rm d}x{\rm d}y}{\int_{\mathbb C}|z|^{2j}\rho_{a}(z){\rm d}x{\rm d}y},\qquad j=0,\dots, d-3.
\]
Again the dependence of the initial conditions $V_j=V_j(a)$, $j\leq d-3$, on the parameter $a$ is controlled by a differential equation for $h=(1,1)$. The initial conditions are
\[
V_0(a)=-\frac{h'(a)}{h(a)},\qquad V_1(a)=-\frac{h''(a)}{h'(a)},\qquad \dots,\qquad
V_{d-3}(a)=-\frac{h^{(d-2)}(a)}{h^{(d-3)}(a)},
\]
with
\begin{equation}\label{e-h}
h(a)=\int_{\mathbb C}\exp\left(-a|z|^2+\frac {\rm i}d\bigl(z^d+\bar z^d\bigr)\right){\rm d}x{\rm d}y.
\end{equation}
As in Lemma \ref{l-1}, we see that this function is a solution of the linear differential equation of order~${d-1}$,
\begin{equation}\label{e-4}
(-1)^{d-1}h^{(d-1)}(a)=ah(a)+a^2h'(a),
\end{equation}
such that $\lim_{a\to \infty}ah(a)=\pi$.

Finally, the rescaled versions (for $d\neq2$)
\smash{$
v_n=a^{{-\frac{2}{d-2}}}V_n
$}
obey the recurrence relations in standard form \cite{AHK2019,CornalbaTaylor1999, Hoppe2025}
\[
v_n+\sum_{j=0}^{d-2}\prod_{k=0}^{d-2}v_{n+j-k}=(n+1)\epsilon, \qquad \epsilon=a^{{-\frac d{d-2}}}.
\]
The initial conditions are $v_{-1}=0$ and
\[
v_j=-a^{{-\frac{2}{d-2}}}\frac{h^{(j+1)}(a)}{h^{(j)}(a)},\qquad j=0,\dots,d-3.
\]

\section{Explicit construction of the orthogonal polynomials}\label{s-4}
The relation \eqref{e-rec1} or the relation \eqref{e-rec2} can be used to determine the orthogonal
polynomials recursively from~${P_j(z)=z^j}$ for $j=0,\dots,d-1$, once we know the $V_n$.
For example, we can write \eqref{e-rec1} as
\begin{equation}\label{e-3tr}
 P_{n+1}(z)=zP_n(z)-\frac {\rm i}a\left(\prod_{j=1}^{d-1}V_{n-j}\right) P_{n-d+1}(z).
\end{equation}
The coefficients are obtained from the recurrence relation \eqref{e-3}. There remains to compute the initial conditions $V_0,\dots,V_{d-3}$. As we have seen, they are determined by the function $h(a)$ given by the integral \eqref{e-h}, solution to the differential equation
\[
(-1)^{d-1}h^{(d-1)}(a)=a h(a)+a^2h'(a)
\]
decaying at infinity. If $d>1$, $a=0$ is a regular point of the differential equation, so the power series method applies. As a result, a basis of local solutions is formed by the generalized hypergeometric functions{\samepage
\begin{equation}\label{e-Bob}
 \phi_m(a)=a^m\cdot {}_1F_{d-3}\left({\frac{m+1}d;
 \frac{m+2}{d},\dots,\frac{d-1}{d},\frac{d+1}{d},\dots, \frac{m+d}{d}};-
 (-d)^{-d+2}a^d\right).
\end{equation}
Here $m$ runs over $0,\dots,d-2$. For $d\geq3$, these series have an infinite radius of convergence.}

To compute the coefficients of $h$ as a linear combination of basis elements, we rewrite the integral as a Laplace transform of the Bessel function $J_0$
\[
 h(a)=\pi\int_0^\infty {\rm e}^{-a t} J_0\left(\frac{2t^\frac d2}{d}\right){\rm d}t.
\]
This formula can be obtained by expanding the exponential of ${\rm i}\bigl(z^d+\bar z^d\bigr)/d$ and performing the angular integration. Now we can expand the exponential series ${\rm e}^{-at}$ and use the identity (see~\cite[formula~(10.22.43)]{NIST:DLMF})
\[
 \int_0^\infty t^\mu J_0(t){\rm d}t=2^\mu
 \frac{\Gamma\left(\frac12(1+\mu)\right)}{\Gamma\left(\frac12(1-\mu)
 \right)}.
\]
We obtain the following statement.

\begin{Proposition} Let $d\geq2$. The function $h$ is given by the linear combination
\[
 h(a)=\pi d^{\frac2d-1}\sum_{m=0}^{d-2}\frac{\bigl(-d^{\frac2d}\bigr)^m}{m!}
 \frac{\Gamma\bigl(\frac{m+1}d\bigr)}
 {\Gamma\bigl(1-\frac{m+1}{d}\bigr)}\phi_m(a)
\]
of the solutions $\phi_0,\dots,\phi_{d-2}$, see \eqref{e-Bob}.
\end{Proposition}
For small $d$, $h$ can be expressed in terms of classical functions.
\begin{itemize}\itemsep=0pt
 \item
 For $d=1$, $a=0$ is not a regular point, not even a regular singular point, of the
 differential equation~\eqref{e-4}, but we can compute the integral explicitly
\[
h(a)=\frac{\pi}a{\rm e}^{-\frac1a}
\]
In this case, $V_n=(n+1)a$ and the orthogonal polynomials are $P_n(z)=\left(z-\frac {\rm i}a\right)^n$,
\item
For $d=2$, the solution is
\[
h(a)=\frac\pi{\sqrt{a^2+1}}.
\]
The recurrence relation \eqref{e-3} reduces to $\big(a^2+1\big)V_n=(n+1)a$ and the recurrence relation~\eqref{e-3tr} for the orthogonal polynomials reduces up to a linear change of variables to the three-term relations for Hermite polynomials $H_{n}(x)=(-1)^n{\rm e}^{x^2}\frac{{\rm d}^n}{{\rm d}x^n}{\rm e}^{-x^2}$. Thus
\[
P_n(z)=\left(\frac\zeta{\sqrt{2(a^2+1)}}\right)^nH_n\left(\zeta^{-1}\sqrt{\frac {a^2+1}2}z\right),\qquad \zeta^2={\rm i}.
\]
Notice that orthogonal polynomials arising in normal matrix models with real Gaussian density, see \cite{Riser2013}, are also related to Hermite polynomials. The new feature here is the rotation of the argument by an eighth root of unity.
\item
For $d=3$, as we have seen,
\[
 h(a)=\sqrt{\frac {\pi a}3} {\rm e}^{a^3/6}K_{\frac16}\bigl(a^3/6\bigr),
\]
and the initial condition leading to the positive solution of \eqref{e-3} is $V_{-1}=0$, $V_{0}=-h'(a)/h(a)$.
\item
 For $d=4$, the third order equation $h'''(a)+ah(a)+a^2h'(a)=0$ reduces with the
 change of variables $h(a)=ay\bigl(a^2/4\bigr)$ to the case $\alpha=1/4$ of the differential equation
 \[
 x^2\frac{{\rm d}^3y}{{\rm d}x^3}+3x\frac{{\rm d}^2y}{{\rm d}x^2}+\bigl(1+4\bigl(x^2-\alpha^2\bigr)\bigr)\frac{{\rm d}y}{{\rm d}x}+4x y=0
 \]
 for products of Bessel functions of order $\alpha$. In fact,
\[
 h(a)=\frac{\pi^2a}4\left(J_{-\frac14}\left({\frac{a^2}4}\right)^2
 -\sqrt2J_{-\frac14}\left({\frac{a^2}4}\right)J_{\frac14}\left({\frac{a^2}4}\right)
 +J_{\frac14}\left({\frac{a^2}4}\right)^2
 \right).
\]
The initial conditions are $V_{-1}=0$, $V_0=-h'(a)/h(a)$, $V_1=-h''(a)/h'(a)$.
\end{itemize}
\section{Relation to quantum minimal surfaces}\label{s-5}
A positive solution of the recurrence relation \eqref{e-3} gives rise to a quantum minimal surface given by operators $W_1$, $W_2$ acting on the space of polynomials. Let
\smash{$
|n\rangle=\frac{P_n}{\sqrt{h_n}}
$}
be the normalized orthogonal polynomials and let \smash{$w_n=\sqrt{v_n}=a^{{-\frac1{d-2}}}\sqrt{V_n}$}. Then the
operator~$L$ and the operator~$M$ of multiplication by $z$ and $L$ can be expressed in terms of the operators $W$, $W^\dagger$ such that
\[
W|n\rangle=w_n|n+1\rangle,\qquad W^{\dagger}|n+1\rangle=w_n|n\rangle \qquad \text{for all}\ n\geq0,
\]
and $W^\dagger|0\rangle=0$.
The relations \eqref{e-rec1}, \eqref{e-rec2} can namely be written as
\begin{align*}
 M = a^{{\frac1{d-2}}}\bigl(W+{\rm i}W^\dagger{}^{d-1}\bigr),
\qquad
 L = a^{{\frac{d-1}{d-2}}}\bigl({\rm i}W^{d-1}+W^\dagger\bigr)
\end{align*}
The relation $[L,M]=1$ between the differential operators $\partial_z+{\rm i}z^{d-1}$ and $z$ translates to
\[
\big[W^\dagger,W\big]+\big[W^\dagger{}^{d-1},W^{d-1}\big]=\epsilon 1, \qquad \epsilon=a^{-\frac{d}{d-2}}.
\]

\section{Conclusion and outlook}\label{s-6}
We have introduced a class of indefinite hermitian inner product on polynomials for which orthogonal polynomials lead to solutions of discrete Painlev\'e-type equations occurring in quantum minimal surfaces. These surfaces are quantizations of the algebraic curves $w_2=w_1^{d-1}$ in~${\mathbb C^2=\mathbb R^4}$. More generally, one expects a similar story to hold for the algebraic curves $w_1^p=w_2^q$ for positive integers $p$, $q$. One may expect that the corresponding orthogonal polynomials are defined by a~density of the form $\rho=|z|^b\exp\bigl(-a|z|^{2p}+{\rm i}t(z^{p+q}+\bar z^{p+q})\bigr)$.

Our results motivate considering the corresponding normal matrix model given by the complex measure
\[
 {\rm d}\mu_N(M)=\exp\bigl(-a\operatorname{tr}\bigl(M^\dagger M\bigr)+{\rm i}\operatorname{tr}\bigl(V(M)+V\bigl(M^\dagger\bigr)\bigr)\bigr){\rm d}M{\rm d}M^\dagger
\]
on normal $N\times N$ matrices for monomial (or more generally entire) potentials $V$. Here $M^\dagger$ denotes the matrix adjoint (conjugate transposed) to $M$ and ${\rm d}M{\rm d}M^\dagger$ is the natural $U(N)$-invariant measure on complex normal matrices, see \cite{ElbauFelder2005,WiegmannZabrodin2000}.
As in the random normal matrix model, the induced complex measure on eigenvalues is
\[
 {\rm d}\mu_N(z)=\exp\left(-a\sum_{j=1}^N|z_j|^2+2{\rm i}\sum_{j=1}^N\operatorname{Re} V(z_j)\right)\prod_{j<k}|z_j-z_k|^2\prod_{j=1}^N
 {\rm d}\operatorname{Re}z_j{\rm d}\operatorname{Im} z_j.
\]
For monomial potential $V$, the partition function $Z_N=\int{\rm d}\mu(M)$ is then up to sign the product of the $h_n$ up to $N$ and is the determinant of the Gram matrix, so Conjecture~\ref{con} is equivalent to the non-vanishing of $Z_N$ for all $N$.
For general polynomial potentials, the complex measure~${\rm d}\mu_N$ is absolutely integrable, as opposed to the real measure
of the normal matrix model of \cite{WiegmannZabrodin2000}, which requires regularization \cite{BleherKuijlaars2012,ElbauFelder2005} to normalize it to a probability measure.
One should expect that, in spite of the fact that the probabilistic interpretation
is not possible, the results of \cite{WiegmannZabrodin2000}, such as the relation to Laplacian growth models and
the description in terms of a real form of the~2D Toda integrable system, hold for the complex measure ${\rm d}\mu_N$.

\subsection*{Acknowledgements} We would like to thank J.~Arnlind, J.~Choe, M.~Duits, P.~Elbau, J.~Fr\"ohlich, A.~Hone, and I.~Kostov for discussions. We thank the anonymous referees for their careful reading and valuable suggestions.

\pdfbookmark[1]{References}{ref}
\LastPageEnding

\end{document}